\documentclass[a4paper,11pt]{article}

\usepackage{amsmath,amssymb,bbm}
\usepackage{cite}
\usepackage[dvipdfm]{graphicx}

\makeatletter

 \@addtoreset{equation}{section}
\makeatother

\newcounter{Enumerate}

\DeclareFontFamily{U}{rsf}{}
\DeclareFontShape{U}{rsf}{m}{n}{
  <5> <6> rsfs5 <7> <8> <9> rsfs7 <10-> rsfs10}{}
\DeclareMathAlphabet\Scr{U}{rsf}{m}{n}

\usepackage[mathscr]{eucal}

\newcommand{\del}{\partial}
\newcommand{\half}{\frac{1}{2}}
\newcommand{\mfk}{\mathfrak}
\newcommand{\LS}{\ \ \ \ \ \ \ \ \ \ }
\newcommand{\ls}{\ \ \ \ \ }
\newcommand{\wt}{\widetilde}

\newcommand{\ol}{\overline}
\newcommand{\dps}{\displaystyle}
\newcommand{\bsubeq}{\begin{subequations}}
\newcommand{\esubeq}{\end{subequations}}
\newcommand{\eps}{\epsilon}
\newcommand{\nn}{\nonumber}

\newcommand{\N}{\mathcal{N}}

\renewcommand{\d}{{\rm d}}
\newcommand{\e}{{\rm e}}
\renewcommand{\i}{{\rm i}}

\renewcommand{\t}{\mfk{t}}
\newcommand{\w}{\wedge}
\newcommand{\slb}{\scalebox}

\renewcommand{\Im}{{\rm Im}}
\renewcommand{\Re}{{\rm Re}}

\begin{document}
\allowdisplaybreaks{

\thispagestyle{empty}


\begin{flushright}
KEK-TH-1481 \\
\end{flushright}

\vspace{30mm}

\begin{center}
\slb{2}{Schwarzschild-AdS Black Holes}

\vspace{5mm}

\slb{2}{in $\N=2$ Geometric Flux Compactification}

\vspace{15mm}

\slb{1.2}{Tetsuji {\sc Kimura}} 

\vspace{2mm}

{\sl
KEK Theory Center,
Institute of Particle and Nuclear Studies, \\
High Energy Accelerator Research Organization \\
Tsukuba, Ibaraki 305-0801, Japan}

\vspace{1mm}

\slb{0.9}{\tt tetsuji\;<at>\;post.kek.jp}

\end{center}

\vspace{10mm}


\begin{abstract}
We present AdS black hole solutions in four-dimensional $\N=2$ gauged supergravity with the universal hypermultiplet.
Here the axion field in this multiplet is dualized to a two-form field.
This system is derived from ten-dimensional massive type IIA theory compactified on nearly-K\"{a}hler manifold in the presence of geometric fluxes and RR-fluxes.
In this work we focus on the simplest coset space $G_2/SU(3)$.
Imposing the covariantly constant condition on all scalar fields, 
we obtain AdS black hole solutions with vanishing electromagnetic charges and arbitrary mass parameter.
\end{abstract}

\newpage

\section{Introduction}
\label{sect-introduction}

Flux compactification is requisite to understand 
low energy effective theories of ten-dimensional theory \cite{Grana:2005jc, Grana:2005ny, Grana:2006hr, D'Auria:2007ay, Cassani:2008rb}.
Calabi-Yau compactification in type II theory yields
four-dimensional $\N=2$ ungauged supergravity.
In this case no scalar potential is involved.
When NSNS three-form flux and RR-fluxes together with non-constant dilaton are turned on in the Calabi-Yau compactification,
non-trivial values of such the fields break the equations of motion in the original ten-dimensional physics.
Hence the modification from Calabi-Yau geometry caused by fluxes has to be considered to derive a genuine effective theory of the ten-dimensional theory.


There are three types of $\N=2$ gauged supergravities via the flux compactifications \cite{D'Auria:2007ay, Cassani:2008rb}:
If there are only the electric flux charge parameters,
the standard gauged supergravity \cite{Andrianopoli:1996cm} emerges.
Once the magnetic RR-flux charge parameters and/or the Romans' mass are involved, 
the axion field of the universal hypermultiplet is dualized to the $B$-field \cite{Dall'Agata:2003yr}.
If the magnetic NSNS-flux charge parameters are also incorporated,
a gauged supergravity with a number of tensor fields \cite{D'Auria:2004yi} is derived.
All of the three types with cubic prepotentials via the flux compactifications are studied in \cite{Cassani:2009na}. 


It is of quite interest to explore black hole solutions in the $\N=2$ gauged supergravities.
Supersymmetric extremal charged AdS black hole solutions with naked singularities \cite{Romans:1991nq, Caldarelli:1998hg, Sabra:1999ux, Chamseddine:2000bk} or with regular event horizons \cite{Cacciatori:2009iz, Dall'Agata:2010gj, Hristov:2010ri} have been well investigated.
There are also developments in the absence of supersymmetry in AdS black hole solutions.
A typical work is \cite{Bellucci:2008cb}, 
where the Fayet-Iliopoulos parameters play an important role in the analysis of the AdS black holes. 
The solutions depend on the symplectic frames.
Notice that both in supersymmetric or in non-supersymmetric cases, AdS black hole solutions with hypermultiplets have not been involved.
On the other hand, in the asymptotically flat case, supersymmetric black hole solution with the universal hypermultiplet is studied \cite{Hristov:2010eu}.
After the Higgs mechanism the system is akin to ungauged supergravity descended from Calabi-Yau compactification with D-branes (see \cite{Ferrara:1997tw, Kallosh:2006bt, Kallosh:2006ib, Nampuri:2007gv} and a lecture note \cite{Bellucci:2006zz}).
However, the gauging necessitates a linear coupling.
This is not realized in the flux compactifications \cite{Louis:2002ny, Cassani:2008rb}.
Gauging the Heisenberg algebra of the hypermultiplets \cite{Sommovigo:2004vj, D'Auria:2004tr} 
is suitable for flux compactifications.


Due to the above story, it is quite intriguing to discover an AdS black hole solution with hypermultiplets, irrespective of preserving supersymmetry, in the framework of flux compactifications.
Here let us introduce the structure of this paper:
In section \ref{sect-GSUGRA} we briefly exhibit $\N=2$ abelian gauged supergravity with $B$-field derived from the compactification on the coset space $G_2/SU(3)$ \cite{KashaniPoor:2007tr, Cassani:2009ck}.
It is known that the four-dimensional $\N=2$ system involves only one vector multiplet and the universal hypermultiplet where the axion field in the hypermultiplet is dualized to the $B$-field \cite{Cassani:2009ck}. 
In section \ref{sect-AdSBH} we analyze black hole solutions in asymptotically AdS spacetime.
In order to reduce complicated interactions we impose the covariantly constant condition on all the scalar fields.
Introducing the static metric ansatz we find AdS black hole solutions.
There we prove that all the electromagnetic charges of the black hole have to vanish caused by the covariantly constant condition.
It turns out that the black hole mass parameter is independent of the expectation values of any scalar fields.
We also argue that the solution is always non-supersymmetric.
In section \ref{sect-discussions} we summarize our results and draw our considerations.
In appendix \ref{app-conventions} we exhibit the convention and the ingredients in the compactification on the coset space $G_2/SU(3)$.


\section{Gauged supergravity with $B$-field}
\label{sect-GSUGRA}

In this section we briefly exhibit the feature of $\N=2$ abelian gauged supergravity with $B$-field derived from massive type IIA compactification on the nearly-K\"{a}hler coset space $G_2/SU(3)$.
The derivations can be seen in \cite{Cassani:2008rb, Cassani:2009na} (see minor differences of the convention in appendix \ref{app-conventions}).

\subsection{Profile of the coset space $G_2/SU(3)$}
\label{subsect-G2SU3}

First of all let us consider the gauged supergravity whose constituents are one vector multiplet and the universal hypermultiplet
associated with the type IIA compactification on $G_2/SU(3)$ \cite{Cassani:2009ck}.
In this compactification,
the moduli space of the vector multiplet is given by $SU(1,1)/U(1)$.
Then the index of the vector fields $A^{\Lambda}$ runs only $\Lambda = 0,1$.
On the other hand, the space of the hypermultiplets is given by $SU(2,1)/U(2)$.
This is expanded only in terms of the scalar fields of the universal hypermultiplet. 
The following flux charge parameters involves the profile of this compactification:
\begin{gather}
e_{10} \ = \ 2 \sqrt{3 \mathscr{I}}
\, , \ \ \ 
m_{\text{R}}^0 \ \neq \ 0
\, , \ \ \ 
e_{\text{R}0} \ \neq \ 0
\, , \label{flux-para-G2-NSR1}
\end{gather}
whilst other flux charges are zero.
Indeed $m_{\text{R}}^0$ is interpreted as the Romans' mass parameter.
The value $\mathscr{I}$ denotes the volume of the coset space.
Notice that the non-vanishing $m_{\text{R}}^0$ makes
the axion field $a$ in the universal hypermultiplet be dualized to the $B$-field \cite{Dall'Agata:2003yr, Cassani:2008rb}.
The moduli space of the vector multiplet, the Hodge-K\"{a}hler geometry,
is governed by the cubic prepotential ${\cal F}(X)$:
\bsubeq \label{TTT-info-NSR1}
\begin{align}
{\cal F} \ &\equiv \ 
\mathscr{I} \frac{X^1 X^1 X^1}{X^0}
\, .
\end{align}
In terms of the local coordinates $\t \equiv X^1/X^0$, we describe the K\"{a}hler potential $K_{\text{V}}$:
\begin{align}
K_{\text{V}} \ &= \ 
- \log \big[ \i (\ol{X}{}^{\Lambda} {\cal F}_{\Lambda} - X^{\Lambda} \ol{\cal F}_{\Lambda}) \big]
\ = \ 
- \log \big[ - \i \mathscr{I} (\t - \ol{\t})^3 \big]
\, . \label{K-metric}
\end{align}
\esubeq

\subsection{Lagrangian and the equations of motion}
\label{EOM-G2SU3}

The Lagrangian associated with the type IIA theory on the coset space $G_2/SU(3)$ \cite{Cassani:2009ck} is 
\bsubeq \label{actionNSR1}
\begin{align}
S \ &= \ 
\int 
\Bigg[
\half R (*1) 
+ \half \mu_{\Lambda \Sigma} (\t, \ol{\t}) F^{\Lambda} \w * F^{\Sigma}
+ \half \nu_{\Lambda \Sigma} (\t, \ol{\t}) F^{\Lambda} \w F^{\Sigma}
- g_{\t \ol{\t}} (\t, \ol{\t}) \d \t \w * \d \ol{\t}
\nn \\
\ & \LS 
- \d \varphi \w * \d \varphi 
- \frac{\e^{- 4 \varphi}}{4} \d B \w * \d B 
- \frac{\e^{2 \varphi}}{2} D \xi^{0} \w * D \xi^{0}
- \frac{\e^{2 \varphi}}{2} D \wt{\xi}_{0} \w * D \wt{\xi}_{0}
- V (*1)
\nn \\
\ & \LS 
+ \half \d B \w \Big\{ 
\xi^0 D \wt{\xi}_0 - \wt{\xi}_0 D \xi^0
+ \big( 2 e_{\text{R} \Lambda} - e_{\Lambda 0} \xi^0 \big) A^{\Lambda}
\Big\}
- \half m^{0}_{\text{R}} e_{\text{R} 0} B \w B
\Bigg]
\, , 
\end{align}
where the functions $\mu_{\Lambda \Sigma} (\t, \ol{\t})$ and $\nu_{\Lambda \Sigma} (\t, \ol{\t})$ are given in (\ref{TTT-periodN-NSR1}).
$g_{\t \ol{\t}}$ is the K\"{a}hler metric defined by (\ref{K-metric}).
The scalar potential $V$ and the covariant derivatives are given as
\begin{gather}
V \ = \ 
g^{\t \ol{\t}} D_{\t} {\cal P}_+ D_{\ol{\t}} \ol{\cal P}_+ 
+ g^{\t \ol{\t}} D_{\t} {\cal P}_3 D_{\ol{\t}} \ol{\cal P}_3 
- 2 |{\cal P}_+|^2 + |{\cal P}_3|^2
\, , \label{V-G2SU3} \\
D \xi^0 \ = \ 
\d \xi^0 
\, , \ls
D \wt{\xi}_0 \ = \ 
\d \wt{\xi}_0 - e_{\Lambda 0} A^0
\, , \label{covderiv-NSR1}
\end{gather}
\esubeq
where the triplet of the Killing prepotentials ${\cal P}_a$ \cite{Cassani:2009na} are explicitly described as
\bsubeq
\begin{gather}
{\cal P}_+ \ = \ 
- 2 \e^{\varphi} L^{1} e_{1 0}
\, , \ls
{\cal P}_- \ = \ 
- 2 \e^{\varphi} L^{1} e_{1 0}
\, , \ls
{\cal P}_3 \ = \ 
\e^{2 \varphi} \big[ L^{0} e_{\text{R}0} - L^{1} e_{1 0} \xi^0 - M_{0} m_{\text{R}}^{0} \big]
\, , \label{KillingPrepot} \\
L^0 \ = \ \e^{K_{\text{V}}/2}
\, , \ls
L^1 \ = \ \t \, \e^{K_{\text{V}}/2}
\, , \ls
M_0 \ = \ 
- \mathscr{I} \, \t^3 \e^{K_{\text{V}}/2}
\, , \\
D_{\t} {\cal P}_a \ = \ 
\Big( \del_{\t} + \half \del_{\t} K_{\text{V}} \Big) {\cal P}_a
\label{killingprepot}
\end{gather}
\esubeq
Due to the absence of the flux charges $e_{\Lambda}{}^0$, the covariant derivative $D\xi^0$ is reduced to the ordinary derivative \cite{Cassani:2008rb}.
Notice that the gauge field strength is defined as $F^{\Lambda} = \d A^{\Lambda} + m_{\text{R}}^{\Lambda} B$ rather than $\d A^{\Lambda}$.
Owing to this, the action (\ref{actionNSR1}) has an additional local symmetry involving the $B$-field such as $\delta B = \d \beta$, 
where the $\beta$ also appears in the variation $\delta A^{\Lambda} = - m_{\text{R}}^{\Lambda} \beta$ with being left the field strength invariant. 
The scalar field $\wt{\xi}_0$ does not contribute to the scalar potential.
The equations of motion in the system (\ref{actionNSR1}) are given as 
\bsubeq \label{eom-NSR1}
\begin{align}
R_{\mu \nu} &- \frac{1}{2} R \, g_{\mu \nu}
\ = \ 
\frac{1}{4} g_{\mu \nu} \, \mu_{\Lambda \Sigma} F^{\Lambda}_{\rho \sigma} F^{\Sigma \rho \sigma} 
- \mu_{\Lambda \Sigma} F^{\Lambda}_{\mu \rho} F^{\Sigma}_{\nu \sigma} \, g^{\rho \sigma}
- g_{\mu \nu} \, g_{\t \ol{\t}} \del_{\rho} \t \del^{\rho} \ol{\t}
+ 2 g_{\t \ol{\t}} \, \del_{\mu} \t \del_{\nu} \ol{\t}
\nn \\
\ & \LS \ls \ \ 
- g_{\mu \nu} \, \del_{\rho} \varphi \del^{\rho} \varphi
+ 2 \del_{\mu} \varphi \del_{\nu} \varphi
- \frac{\e^{-4 \varphi}}{24} g_{\mu \nu} \, H_{\rho \sigma \lambda} H^{\rho \sigma \lambda} 
+ \frac{\e^{-4 \varphi}}{4} H_{\mu \rho \sigma} H_{\nu \lambda \delta} \, g^{\rho \lambda} g^{\sigma \delta}
\nn \\
\ & \LS \ls \ \ 
- \frac{\e^{2 \varphi}}{2} g_{\mu \nu} \Big( D_{\rho} \xi^{0} D^{\rho} \xi^{0}
+ D_{\rho} \wt{\xi}_0 D^{\rho} \wt{\xi}_0 \Big)
+ \e^{2 \varphi} \Big( D_{\mu} \xi^{0} D_{\nu} \xi^{0} 
+ D_{\mu} \wt{\xi}_0 D_{\nu} \wt{\xi}_0 \Big)
- g_{\mu \nu} V
\, , \label{eom_g_NSR1} \\
0 \ &= \ 
- \frac{1}{\sqrt{-g}} \del_{\mu} \Big( \sqrt{-g} (* \wt{F}_{\Lambda})^{\mu \sigma} \Big)
+ \frac{\eps^{\mu \nu \rho \sigma}}{2 \sqrt{-g}} \del_{\mu} B_{\nu \rho} 
\big( e_{\text{R}\Lambda} - e_{\Lambda 0} \xi^0 \big)
- \e^{2 \varphi} \, e_{\Lambda 0} D^{\sigma} \wt{\xi}_0
\, , \label{eom_A_NSR1} \\
0 \ &= \ 
\frac{1}{\sqrt{-g}} \del_{\mu} \Big( \sqrt{-g} \, g_{\t \ol{\t}} \, g^{\mu \nu} \del_{\nu} \ol{\t} \Big) 
+ \frac{1}{4} \frac{\del \mu_{\Lambda \Sigma}}{\del \t} F^{\Lambda}_{\mu \nu} F^{\Sigma \mu \nu}
- \frac{\eps^{\mu \nu \rho \sigma}}{8 \sqrt{-g}} 
\frac{\del \nu_{\Lambda \Sigma}}{\del \t} F^{\Lambda}_{\mu \nu} F^{\Sigma}_{\rho \sigma}
\nn \\
\ & \ \ \ \
- \del_{\t} g_{\t \ol{\t}} \, \del_{\mu} \t \del^{\mu} \ol{\t}
- \frac{\del V}{\del \t}
\, , \label{eom_t_NSR1} \\
0 \ &= \ 
\frac{2}{\sqrt{-g}} \del_{\mu} \Big( \sqrt{-g} \, g^{\mu \nu} \del_{\nu} \varphi \Big) 
+ \frac{\e^{4 \varphi}}{6} H_{\mu \nu \rho} H^{\mu \nu \rho}
- \e^{2 \varphi} \Big( 
D_{\mu} \xi^{0} D^{\mu} \xi^{0} + D_{\mu} \wt{\xi}_0 D^{\mu} \wt{\xi}_0 \Big)
- \frac{\del V}{\del \varphi}
\, , \label{eom_dilaton_NSR1} \\
0 \ &= \ 
\frac{1}{\sqrt{-g}} \del_{\mu} \Big( \e^{-4 \varphi} \sqrt{-g} H^{\mu \rho \sigma} \Big)
+ \frac{\eps^{\mu \nu \rho \sigma}}{\sqrt{-g}} \Big[
2 D_{\mu} \xi^0 D_{\nu} \wt{\xi}_0
+ \big( e_{\text{R}\Lambda} - e_{\Lambda 0} \xi^0 \big) F^{\Lambda}_{\mu \nu}
\Big]
\nn \\
\ & \ \ \ \ 
+ 2 m_{\text{R}}^{0} \, \mu_{0 \Sigma} \, F^{\Sigma \rho \sigma} 
- \frac{\eps^{\mu \nu \rho \sigma}}{\sqrt{-g}} m_{\text{R}}^{0} \,
\nu_{0 \Sigma} \, F^{\Sigma}_{\mu \nu} 
\, , \label{eom_B_NSR1} \\
0 \ &= \ 
- \frac{2}{\sqrt{-g}} \del_{\mu} \Big(
\sqrt{-g} \, \e^{2 \varphi} \, g^{\mu \nu} D_{\nu} \xi^{0} \Big) 
+ \frac{\del V}{\del \xi^{0}}
+ \frac{\eps^{\mu \nu \rho \sigma}}{2\sqrt{-g}} \del_{\mu} B_{\nu \rho} 
D_{\sigma} \wt{\xi}_0
\, , \label{eom_xi_NSR1} \\
0 \ &= \ 
- \frac{2}{\sqrt{-g}} \del_{\mu} \Big(
\sqrt{-g} \, \e^{2 \varphi} \, g^{\mu \nu} D_{\nu} \wt{\xi}_0 \Big) 
- \frac{\eps^{\mu \nu \rho \sigma}}{2\sqrt{-g}} \del_{\mu} B_{\nu \rho} 
D_{\sigma} \xi^{0} 
\, . \label{eom_wtxi_NSR1}
\end{align}
\esubeq
The field strength of the $B$-field is given as $H_{\mu \nu \rho} = 3 \del_{[\mu} B_{\nu \rho]}$.
It is worth introducing a dual tensor of the gauge field strength $F^{\Lambda}_{\mu \nu}$ in order to define electromagnetic charges \cite{Cassani:2008rb}:
\bsubeq
\begin{align}
\wt{F}_{\Lambda \mu \nu} \ &\equiv \ 
\nu_{\Lambda \Sigma} F^{\Sigma}_{\mu \nu}
+ \mu_{\Lambda \Sigma} (*F^{\Sigma})_{\mu \nu}
\, . \label{dual-F-NSR1}
\end{align}
The Hodge dual in the Lagrangian is defined in terms of the metric and a constant tensor $\eps_{\mu \nu \rho \sigma}$:
\begin{gather}
(*F^{\Lambda})_{\mu \nu} \ \equiv \ 
\frac{\sqrt{-g}}{2} \eps_{\mu \nu \rho \sigma} F^{\Lambda \rho \sigma}
\, .
\end{gather}
\esubeq
We normalize the constant tensor as $\eps_{0123} = + 1$ and its contravariant tensor as $\eps^{0123} = - 1$ in a generic curved spacetime. 

If all the flux charges were zero,
the internal space $G_2/SU(3)$ would be reduced to a Calabi-Yau three-fold.
In this situation the covariant derivatives would also be reduced to the ordinary derivatives and the scalar potential would become trivially zero.
Hence the system would be reduced to an ungauged supergravity.


\section{AdS black hole solutions}
\label{sect-AdSBH}

In this section we analyze black hole solutions.
It is tough to solve the equations of motion (\ref{eom-NSR1}) without appropriate ans\"{a}tze.
In order to facilitate solving the equations of motion in a gauge covariant way,
we introduce the covariantly constant condition.
Furthermore, we focus on extremal, static, possibly charged black holes.
Thus the static metric ansatz and electromagnetic charges are incorporated.
We solve the Einstein equation under the static metric ansatz.
We recognize that scalar fields are constant under the covariantly constant condition.
Finally we investigate possible values of the black hole charges.


\subsection{Covariantly constant solution}
\label{subsect-covconst}

Here we consider an appropriate condition to solve the equations of motion (\ref{eom-NSR1}).
The simplest condition is the constant condition imposed on all the fields.
However, owing to the form of the covariant derivatives (\ref{covderiv-NSR1}), 
the constant condition breaks the gauge invariance of the equations of motion.
In order to preserve the gauge invariance, we have to introduce an alternative condition.
Since all the terms in (\ref{eom-NSR1}) are given in the gauge covariant way,
the covariantly constant condition seems to be a suitable condition.
As illustrated later, 
this condition indeed plays a powerful role in the analysis.

The covariantly constant condition is introduced as follows:
\bsubeq \label{covconst-NSR1}
\begin{align}
\del_{\mu} \t \ &= \ 0
\, , \ \ \ 
\del_{\mu} \varphi \ = \ 0
\, , \ \ \ 
D_{\mu} \xi^0 \ = \ 0
\, , \ \ \ 
D_{\mu} \wt{\xi}_0 \ = \ 0
\, . \label{covconst-scalar}
\end{align}
To simplify the equations, we also impose that the $B$-field is closed:
\begin{align}
\del_{[\mu} B_{\nu \rho]} \ &= \ 0
\, . \label{B-closed}
\end{align}
\esubeq
The field equation for the gauge fields is reminiscent of the one in ungauged supergravity derived from Calabi-Yau compactification,
whilst the field equations for the scalar fields and the $B$-field remain non-trivial.


\subsection{Metric ansatz and electromagnetic charges}
\label{sect-static-ansatz-NSR1}

Next we introduce a metric ansatz.
Since we would like to find an AdS black hole solution, 
we have to introduce the asymptotically AdS spacetime metric.
For simplicity we focus only on the extremal, static, spherically symmetric black hole whose metric can be given as
\begin{align}
\d s^2 \ &= \ 
- \e^{2 A(r)} \d t^2 + \e^{- 2 A(r)} \d r^2 
+ \e^{2 C(r)} r^2 \, \big( \d \theta^2 + \sin^2 \theta \, \d \phi^2 \big) 
\, . \label{static-NSR1}
\end{align}

We define electromagnetic charges $(p^{\Lambda}, q_{\Lambda})$ of the system in terms of the gauge field strengths:
\begin{align}
p^{\Lambda} \ &\equiv \ \frac{1}{4 \pi} \int_{S^2} F^{\Lambda}_2
\, , \ls
q_{\Lambda} \ \equiv \ \frac{1}{4 \pi} \int_{S^2} \wt{F}_{\Lambda 2}
\, . \label{BH-charge-NSR1}
\end{align}
Since all the interaction terms in (\ref{eom_A_NSR1}) vanish because of the covariantly constant condition (\ref{covconst-NSR1}), 
each component of the gauge field strength can be easily evaluated: 
\bsubeq \label{pq-F}
\begin{align}
F^{\Lambda}_{\theta \phi} \ &= \ 
p^{\Lambda} \sin \theta
\, , \\
\wt{F}_{\Lambda \theta \phi} \ &= \ 
q_{\Lambda} \sin \theta
\ = \ 
\nu_{\Lambda \Sigma} F^{\Sigma}_{\theta \phi} 
+ \mu_{\Lambda \Sigma} \Big( \sqrt{-g} \, \eps_{\theta \phi t r} F^{\Sigma t r} \Big)
\, , \\
F^{\Lambda}_{t r} \ &= \ 
- \frac{\e^{-2 C}}{r^2} (\mu^{-1})^{\Lambda \Sigma} \big( q_{\Sigma} - \nu_{\Sigma \Gamma} p^{\Gamma} \big)
\, .
\end{align}
\esubeq
The energy momentum tensor of the gauge fields in the Einstein equation (\ref{eom_g_NSR1}) is evaluated:
\bsubeq \label{sympl-inv-covconst-NSR1} 
\begin{align}
T_{\mu}{}^{\nu} \ &= \ 
T_{\mu \rho} \, g^{\rho \nu} 
\ \equiv \ 
\Big[
\frac{1}{4} g_{\mu \rho} \, \mu_{\Lambda \Sigma} \, F^{\Lambda}_{\lambda \sigma} F^{\Sigma \lambda \sigma} 
- \mu_{\Lambda \Sigma} \, F^{\Lambda}_{\mu \lambda} F^{\Sigma}_{\rho \sigma} \, g^{\lambda \sigma} 
\Big]
g^{\rho \nu} 
\, . \label{stress-covconst-NSR1} 
\end{align}
By virtue of the description (\ref{pq-F}),
the energy momentum tensor is recast as the first symplectic invariant:
\begin{align}
T_t{}^t \ &= \ 
T_r{}^r \ = \ 
- T_{\theta}{}^{\theta} \ = \ 
- T_{\phi}{}^{\phi} \ = \ 
- \frac{\e^{-4 C}}{r^4} I_1
\, , \\
I_1 (p,q) \ &\equiv \ 
- \half \Big[
p^{\Lambda} \mu_{\Lambda \Sigma} p^{\Sigma} 
+ (q_{\Lambda} - \nu_{\Lambda \Gamma} p^{\Gamma}) (\mu^{-1})^{\Lambda \Sigma}
(q_{\Sigma} - \nu_{\Sigma \Delta} p^{\Delta})
\Big]
\, . 
\end{align}
The symplectic invariant $I_1$ also appears 
in the equation of motion for the vector modulus $\t$ (\ref{eom_t_NSR1}):
\begin{align}
\frac{1}{4} \frac{\del \mu_{\Lambda \Sigma}}{\del \t} 
F^{\Lambda}_{\mu \nu} F^{\Sigma \mu \nu}
- \frac{\eps^{\mu \nu \rho \sigma}}{8 \sqrt{-g}} 
\frac{\del \nu_{\Lambda \Sigma}}{\del \t}
F^{\Lambda}_{\mu \nu} F^{\Sigma}_{\rho \sigma}
\ &= \ 
- \frac{\e^{- 4 C}}{r^4} \frac{\del I_1}{\del \t}
\, . \label{del-I_1-covconst-NSR1}
\end{align}
\esubeq
Utilizing (\ref{covconst-NSR1}), (\ref{static-NSR1}) and (\ref{sympl-inv-covconst-NSR1}),
we reduce the equations of motion (\ref{eom-NSR1}) to
\bsubeq \label{eom-NSR1-reduce2}
\begin{align}
0 \ &= \ 
\e^{2 A} \Big[
\frac{1}{r^2} (1 - \e^{-2(A+C)}) 
+ \frac{2}{r} (A'+ 3 C') 
+ C' (2 A' + 3C') 
+ 2 C''
\Big]
+ \frac{\e^{-4C}}{r^4} I_1 + V
\, , \label{eom_g00-covconst1} \\
0 \ &= \ 
\e^{2 A} \Big[
\frac{1}{r^2} (1 - \e^{-2(A+C)}) + \frac{2}{r}(A' + C') + C'(2A' + C') 
\Big]
+ \frac{\e^{-4C}}{r^4} I_1 + V
\, , \label{eom_g11-covconst1} \\
0 \ &= \ 
\e^{2 A} \Big[
\frac{2}{r} (A' + C') + 2 (A')^2 + C'(2A' + C') + A'' + C'' 
\Big]
- \frac{\e^{-4C}}{r^4} I_1 + V
\, , \label{eom_g22-covconst1} \\
0 \ &= \ 
\frac{\e^{-4C}}{r^4} \frac{\del I_1}{\del \t} 
+ \frac{\del V}{\del \t}
\, , \label{eom_t-covconst1} \\
0 \ &= \ 
\frac{\del V}{\del \varphi} 
\ = \
2 V_{\text{NS}} + 4 V_{\text{R}}
\, , \label{eom_dilaton-covconst1} \\
0 \ &= \ 
m_{\text{R}}^{0} \mu_{0 \Sigma} F^{\Sigma \rho \sigma} 
- \frac{\eps^{\mu \nu \rho \sigma}}{2 \sqrt{-g}} 
m_{\text{R}}^{0} \nu_{0 \Sigma} F^{\Sigma}_{\mu \nu} 
+ e_{\text{R} 0} 
\frac{\eps^{\mu \nu \rho \sigma}}{2 \sqrt{-g}} F^{0}_{\mu \nu} 
- e_{1 0} \xi^0 
\frac{\eps^{\mu \nu \rho \sigma}}{2 \sqrt{-g}} F^{1}_{\mu \nu} 
\, , \label{eom_B-covconst1} \\
0 \ &= \ 
\frac{\del V}{\del \xi^0}
\ = \ 
\e^{4 \varphi} e_{1 0} (\mu^{-1})^{1 \Sigma}
\Big[ (e_{\text{R}\Sigma} - e_{\Sigma 0} \xi^0) - \nu_{\Sigma 0} m_{\text{R}}^{0}
\Big]
\, . \label{eom_xi-covconst1} 
\end{align}
\esubeq
Here the prime denotes the derivative with respect to the radial coordinate $r$.
The scalar potentials $V_{\text{NS}}$ and $V_{\text{R}}$ in (\ref{eom_dilaton-covconst1}) are defined in \cite{Cassani:2009na}.


\subsection{Evaluation of the metric}
\label{subsect-eval-metric}

Next task is to solve the equations (\ref{eom-NSR1-reduce2}).
The difference between 
(\ref{eom_g00-covconst1}) and (\ref{eom_g11-covconst1}) 
gives rise to the differential equation for the scale function $C(r)$:
\bsubeq
\begin{align}
0 \ &= \ 
\frac{2}{r} C' + (C')^2 + C'' 
\, . \label{derivative-C-covconst1}
\end{align} 
The solution has two integration constants $c_1$ and $c_2$: 
\begin{align}
C(r) \ &= \ 
c_2 + \log \Big( c_1 + \frac{1}{r} \Big) 
\, . \label{sol-C-covconst-NSR1}
\end{align}
\esubeq
Substituting this into the equation
(\ref{eom_g22-covconst1}), we obtain the differential equation for the scale function $A(r)$. 
Its solution is given in terms of integration constants $a_1$ and $a_2$ as follows:
\bsubeq
\begin{align}
\e^{2 A(r)} \ &= \ 
\e^{-4c_2}\frac{6 I_1 - \e^{4c_2} (c_1r + 1)}{3c_1^2 (c_1 r + 1)^2}
\Big[ (c_1r + 1)^3 V + 6 c_1 \big\{ a_1 - c_1 a_2 (c_1 r + 1) \big\} \Big]
\, . \label{presol-A-covconst-NSR1}
\end{align}
The equation (\ref{eom_g00-covconst1}) yields the relation among three integration constants in (\ref{sol-C-covconst-NSR1}) and (\ref{presol-A-covconst-NSR1}):
\begin{align}
a_2 \ &= \ \frac{\e^{- 2 c_2}}{2 (c_1)^2}
\, . \label{constraint-ab-covconst-NSR1}
\end{align}
This implies that the integration constants are restricted such that 
$c_1$ is non-zero, $a_2$ is positive and $c_2$ is finite. 
Substituting (\ref{constraint-ab-covconst-NSR1}) into (\ref{presol-A-covconst-NSR1}), we obtain a familiar form: 
\begin{align}
\e^{2 A(r)} \ &= \ 
\frac{\e^{-2c_2}}{(c_1)^2} - \frac{2 a_1}{c_1 (c_1 r + 1)} 
+ \frac{\e^{- 4 c_2} I_1}{(c_1)^2 (c_1 r + 1)^2}
- \frac{V}{3 (c_1)^2} (c_1 r + 1)^2 
\, . \label{sol-A-covconst-NSR1}
\end{align}
\esubeq
Without loss of generality, we can fix the integration constants 
$c_1 = 1$ and $c_2 = 0$. It is worth re-defining the radial coordinate as $\wt{r} \equiv r + 1$.
Substituting (\ref{sol-C-covconst-NSR1}) and (\ref{sol-A-covconst-NSR1}) into (\ref{static-NSR1}), we rewrite the line element of the four-dimensional spacetime:
\bsubeq \label{metric-sol-covconst-NSR1}
\begin{align}
\d s^2 \ &= \ 
- V(\wt{r}) \d t^2 
+ \frac{1}{V(\wt{r})} \d \wt{r}{}^2 
+ \wt{r}{}^2 \big( 
\d \theta^2 + \sin^2 \theta \, \d \phi^2 \big)
\, , \\
V(\wt{r}) \ &= \ 
1 - \frac{2 a_1}{\wt{r}} 
+ \frac{I_1}{\wt{r}{}^2}
- \frac{V}{3} \wt{r}{}^2
\, .
\end{align}
\esubeq
We can read the various parameters in $V(\wt{r})$: the first term in the right-hand side represents the scalar curvature of the horizon; $a_1$, $I_1$ and $V$ are interpreted as the black hole mass parameter $\eta$, the square of the black hole charges, and the cosmological constant $\Lambda$, respectively.


\subsection{Evaluation of the matter fields and the scalar potential}
\label{subsect-scalars}

Our next task is to analyze the equations of motion for matter fields
(\ref{eom_t-covconst1}), (\ref{eom_dilaton-covconst1}), and (\ref{eom_xi-covconst1}).
Since we imposed the covariantly constant condition (\ref{covconst-NSR1}),
the vector modulus $\t$ does not depend on any spacetime coordinates.
Then the symplectic invariant $I_1$ is also independent of the spacetime coordinates.
This indicates that the derivatives $\del I_1/\del \t$ and $\del V/\del \t$ should be zero independently\footnote{If the equation $\del V/ \del \t = 0$ does not exist, the equation $\del I_1/ \del \t = 0$ is nothing but the one in search of extremal black holes in ungauged supergravity 
(see \cite{Kallosh:2006ib, Nampuri:2007gv, Bellucci:2006zz} and references therein).}:
\begin{align}
\frac{\del I_1}{\del \t} \ &= \ 0 \ = \ 
\frac{\del V}{\del \t}
\, . \label{delI-delV-covconst-NSR1}
\end{align}
Apart from the evaluation of $\del I_1/ \del \t = 0$,
the other equations correspond to the field equations of the vacua. 
Hence, as illustrated in (\ref{sol-NSR1-vacua1}), (\ref{sol-NSR1-vacua2}) and (\ref{sol-NSR1-vacua3}), the scalar fields and the scalar potential $(\varphi, \xi^0, V)$ except for $\wt{\xi}_0$ are determined:
\begin{align}
\{ \t, \xi^0 , \varphi , V \}_{\text{black hole}}
\ &= \ 
\left\{
\begin{array}{l}
\{ t_*, \xi^0_* , \varphi_* , V_* \} \Big|_{\text{(\ref{sol-NSR1-vacua1})}}
\, , \\
\{ t_*, \xi^0_* , \varphi_* , V_* \} \Big|_{\text{(\ref{sol-NSR1-vacua2})}}
\, , \\
\{ t_*, \xi^0_* , \varphi_* , V_* \} \Big|_{\text{(\ref{sol-NSR1-vacua3})}}
\, .
\end{array}
\right.
\label{sol-scalars-covconst1}
\end{align}


\subsection{Black hole charges}
\label{subsect-BH-charges}

Here we analyze the black hole charges $(p^{\Lambda}, q_{\Lambda})$ defined in (\ref{BH-charge-NSR1}). 
They satisfy the equations of motion for $\t$, $B_{\mu \nu}$ and the covariantly constant condition for $\wt{\xi}_0$.
They are described as follows:
\bsubeq
\begin{align}
0 \ &= \ 
\frac{\del I_1}{\del \t} 
\, , \label{delI_t-covconst1} \\
0 \ &= \ 
m_{\text{R}}^0 \mu_{0 \Sigma} \Big( \frac{\sqrt{-g}}{2} \eps_{\lambda \gamma \rho \sigma} F^{\Sigma \rho \sigma} \Big)
+ m_{\text{R}}^0 \nu_{0 \Sigma} F^{\Sigma}_{\lambda \gamma}
- e_{\text{R}0} F^0_{\lambda \gamma}
+ e_{10} \xi^0 F^1_{\lambda \gamma}
\, , \label{eom_B-covconst1-2} \\
0 \ &= \ 
D_{\mu} \wt{\xi}_0 \ = \ 
\del_{\mu} \wt{\xi}_0 - e_{\Lambda 0} A^{\Lambda}_{\mu}
\, . \label{wtxi-covconst1}
\end{align}
\esubeq

First let us focus on (\ref{wtxi-covconst1}). 
Performing the commutator of the ordinary derivative acting on $\wt{\xi}_0$, we obtain a non-trivial equation for the field strength:
\begin{align}
0 \ &= \ [ \del_{\mu}, \del_{\nu} ] \wt{\xi}_0 \ = \ e_{\Lambda 0} F^{\Lambda}_{\mu \nu}
\, .
\end{align}
This equation implies that $F^1_{\mu \nu}$ vanishes because the flux charge $e_{10}$ is non-zero. 
Applying this to (\ref{pq-F}), we obtain the following two equations:
\bsubeq \label{zero-F1}
\begin{align}
0 \ &= \ F^1_{\theta \phi} \ = \ 
p^1 \sin \theta
\, , \\
0 \ &= \ F^1_{t r} \ = \ 
- \frac{\e^{- 2C(r)}}{r^2} (\mu^{-1})^{1 \Sigma} (q_{\Sigma} - \nu_{\Sigma \Gamma} p^{\Gamma})
\, .
\end{align}
\esubeq
These two denote the condition among the charges:
\bsubeq \label{charge-condition1-covconst1}
\begin{align}
0 \ &= \ p^1
\, , \label{p1-covconst1} \\
0 \ &= \ 
(\mu^{-1})^{10} \, q_0 + (\mu^{-1})^{11} \, q_1 - (\mu^{-1} \nu)^1{}_0 \, p^0
\, . \label{q0-q1-p0-covconst1}
\end{align}
\esubeq
Owing to (\ref{zero-F1}), the field equation (\ref{eom_B-covconst1-2}) is reduced to
\begin{align}
e_{\text{R}0} \, F^0_{\lambda \gamma}
\ &= \ 
m_{\text{R}}^0 \, \nu_{0 \Sigma} \, F^{\Sigma}_{\lambda \gamma}
+ m_{\text{R}}^0 \, \mu_{0 \Sigma} \Big( \frac{\sqrt{-g}}{2} \eps_{\lambda \gamma \rho \sigma} F^{\Sigma \rho \sigma} \Big)
\, , 
\end{align}
which generates two following relations:
\bsubeq
\begin{align}
p^0 \ &= \ 
\frac{m_{\text{R}}^0}{e_{\text{R}0}} q_0
\, , \label{p0-q0-covconst1} \\
0 \ &= \ 
p^{\Sigma} \Big[
m_{\text{R}}^0 \, \mu_{0 \Sigma} 
+ m_{\text{R}}^0 (\nu \mu^{-1} \nu)_{0 \Sigma}
- e_{\text{R}0} (\mu^{-1} \nu)^0{}_{\Sigma}
\Big]
\nn \\
\ & \ \ \ \
- q_{\Sigma} \Big[ 
m_{\text{R}}^0 (\nu \mu^{-1})_0{}^{\Sigma} - e_{\text{R}0} (\mu^{-1})^{0 \Sigma}
\Big]
\, . \label{q-p-covconst1}
\end{align}
\esubeq
Due to (\ref{p0-q0-covconst1}),
the charge $q_0$ is related to the other electric charge $q_1$ via (\ref{q0-q1-p0-covconst1}):
\begin{align}
(\mu^{-1})^{11} \, q_1 \ &= \ 
- q_0 \Big[ (\mu^{-1})^{10} - \Big( \frac{m_{\text{R}}^0}{e_{\text{R}0}} \Big) (\mu^{-1} \nu)^1{}_0 
\Big]
\, . \label{q1-q0-covconst1}
\end{align}
Furthermore, substituting (\ref{p0-q0-covconst1}) and (\ref{q1-q0-covconst1}) into (\ref{q-p-covconst1}) and multiplying $(\mu^{-1})^{11}/e_{\text{R}0}$, we find an equation for the charge $q_0$:
\begin{align}
0 \ &= \ 
q_0 \Big[
\Big( (\mu^{-1})^{11} (\mu^{-1})^{00} - [(\mu^{-1})^{01}]^2 \Big)
- 2 \Big( \frac{m_{\text{R}}^0}{e_{\text{R}0}} \Big)
\Big( (\mu^{-1})^{11} (\mu^{-1} \nu)^0{}_0 - (\mu^{-1})^{01} (\mu^{-1} \nu)^1{}_0 \Big)
\nn \\
\ &\LS
+ \Big( \frac{m_{\text{R}}^0}{e_{\text{R}0}} \Big)^2
\Big( (\mu^{-1})^{11} (\mu + \nu \mu^{-1} \nu)_{00} 
- [(\mu^{-1} \nu)^1{}_0 ]^2 \Big)
\Big]
\, , \label{q0-covconst1}
\end{align}
where we used symmetry on the period matrix $(\mu^{-1} \nu)^{\Lambda}{}_{\Sigma} = (\nu \mu^{-1})_{\Sigma}{}^{\Lambda}$ and $(\mu^{-1})^{\Lambda \Sigma} = (\mu^{-1})^{\Sigma \Lambda}$.
In terms of the explicit forms (\ref{TTT-periodN-NSR1}) with $\t \equiv x + \i y$, 
we obtain the followings: 
\bsubeq
\begin{align}
(\mu^{-1})^{11} (\mu^{-1})^{00} - [(\mu^{-1})^{01}]^2 
\ &= \ 
\frac{1}{3 \mathscr{I} y^4}
\, , \\
(\mu^{-1})^{11} (\mu^{-1} \nu)^0{}_0 - (\mu^{-1})^{01} (\mu^{-1} \nu)^1{}_0 
\ &= \ 
\frac{2 x^3}{3 \mathscr{I} y^4}
\, , \\
(\mu^{-1})^{11} (\mu + \nu \mu^{-1} \nu)_{00} - [(\mu^{-1} \nu)^1{}_0 ]^2 
\ &= \ 
\frac{(x^2 + y^2)^2 (4 x^2 + y^2)}{3 y^4}
\, .
\end{align}
\esubeq
Substituting them into (\ref{q0-covconst1}), we evaluate the value of $q_0$ on each solution of (\ref{sol-scalars-covconst1}):
\bsubeq
\begin{alignat}{2}
0 \ &= \ \frac{72}{25} q_0
\left( 
\frac{24 (m_{\text{R}}^0)^2}{5 (e_{10} e_{\text{R}0})^2}
\right)^{2/3} 
&\ls \ \ & \text{at (\ref{sol-NSR1-vacua1})}
\, , \\
0 \ &= \ 
4 q_0 
\left( 
\frac{24 (m_{\text{R}}^0)^2}{5 (e_{10} e_{\text{R}0})^2}
\right)^{2/3}
&\ls \ \ & \text{at (\ref{sol-NSR1-vacua2})}
\, , \\
0 \ &= \ 
2 q_0 
\left( 
\frac{12 (m_{\text{R}}^0)^2}{\sqrt{5} (e_{10} e_{\text{R}0})^2}
\right)^{2/3}
&\ls \ \ & \text{at (\ref{sol-NSR1-vacua3})}
\, .
\end{alignat}
\esubeq
This indicates that the black hole charge $q_0$ must be zero because any flux charge parameters $(e_{10}, e_{\text{R}0}, m_{\text{R}}^0)$ are non-zero.
This result with (\ref{p1-covconst1}), (\ref{p0-q0-covconst1}) and (\ref{q1-q0-covconst1}) makes all the black hole charges vanish (\ref{sol-scalars-covconst1}):
\begin{align}
q_0 \ &= \ 0 
\, , \ \ \ 
q_1 \ = \ 0
\, , \ \ \ 
p^0 \ = \ 0 
\, , \ \ \ 
p^1 \ = \ 0
\, . \label{sol-BH-charges-NSR1}
\end{align}
This configuration automatically satisfies the equation (\ref{delI_t-covconst1}).

Here let us elucidate the configuration of the gauge potentials and the $B$-field under the covariantly constant condition (\ref{covconst-NSR1}).
Since $m_{\text{R}}^1$ and $F^1_{\mu \nu}$ vanish, 
the gauge potential $A^1_{\mu}$ is given as $A^1_{\mu} = \del_{\mu} \alpha^1$, where $\alpha^1$ is arbitrary. 
The vanishing $F^0_{\mu \nu}$ illustrates the relation between
the graviphoton $A^0_{\mu}$ and the $B$-field in such a way as $2 \del_{[\mu} A^0_{\nu]} = - m_{\text{R}}^0 B_{\mu \nu}$.
This indicates that the $B$-field and the graviphoton are described as
$B_{\mu \nu} = 2 \del_{[\mu} \beta_{\nu]}$ and $A^0_{\mu} = \del_{\mu} \alpha^0 - m_{\text{R}}^0 \beta_{\mu}$, where $\beta_{\mu}$ and $\alpha^0$ are arbitrary.
Indeed the values $\alpha^{\Lambda}$ and $\beta_{\mu}$ are parameters of gauge transformations and the local symmetry of the $B$-field, respectively.

The BPS bound equation $M^2 = |{\cal Z}|^2$ is useful to consider if a solution is supersymmetric or not.
Here $M$ is the black hole physical mass corresponding to the black hole mass parameter $\eta = a_1$ if topology of the horizon is two-sphere (see, for instance, \cite{Caldarelli:1998hg}).
Since each solution has no black hole charges,
the BPS equation is satisfied if and only if the mass parameter vanishes.
This represents the AdS vacuum itself.
Hence our black hole solutions with the non-vanishing mass parameter are non-supersymmetric.


\section{Summary and discussions}
\label{sect-discussions}

In this paper we studied $\N=2$ abelian gauged supergravity with $B$-field via geometric flux compactification of type IIA theory.
First we illustrated the profile of the compactification on the nearly-K\"{a}hler coset space $G_2/SU(3)$.
Next, 
we introduced the covariantly constant condition on all the scalar fields.
This simplifies the equations of motion for all the bosonic fields.
We further restricted ourselves to study the extremal, static, spherically symmetric black holes in the asymptotically AdS spacetime.
It turns out that the scalar fields $(\t, \xi^0, \varphi)$ remain constant,
whilst the other scalar field $\wt{\xi}_0$ is arbitrary.
The value of the scalar potential $V$ is interpreted as the cosmological constant.
The symplectic invariant $I_1$ is regarded as the square of the black hole charges.
It also turns out that the covariantly constant condition forces the charges to be zero.
In addition, we found that the integration constant $a_1$ behaves as the black hole mass parameter.
This value is not affected by any flux charges and gauge fields.
Eventually
we obtained Schwarzschild-AdS black hole solutions with arbitrary mass parameter.
We recognized that the black hole solutions are always non-supersymmetric irrespective of the supersymmetry of the vacua,
because the BPS equation is satisfied if and only if the mass parameter vanishes.
In the absence of the magnetic RR-flux charge parameters except for the Romans' mass, 
the gauge field $A^1_{\mu}$ in the vector multiplet is trivial up to the gauge transformation.
On the other hand, the graviphoton $A^0_{\mu}$ and the $B$-field are closely related to each other via the Romans' mass parameter 
under the gauge transformation and the local symmetry of the $B$-field.
Since they are also asymptotically connected to the values in the vacuum,
their values in the bulk are also trivial up to the gauge transformations.
This phenomenon is different from the one in ungauged supergravity derived from a Calabi-Yau compactification with D-branes.
In that case one can find an extremal black hole solution with constant scalar fields. 
There all the physical values such as the mass parameter and the constant scalars are determined by the D-brane charges.

This work convinces us that non-constant fields will be necessary to build a charged AdS black hole solution.
More precisely, the covariantly constant condition (\ref{covconst-NSR1}) must be relaxed in order to see a charged AdS black hole.
The electromagnetic charges are defined by the gauge field strength descended from RR-fluxes.
In our analysis, the dynamics of charged particles are turned off caused by the covariantly constant condition.
Technically it is difficult to relax the covariantly constant condition for the whole fields (\ref{covconst-NSR1}),
whilst it might be interesting if a part of the condition is relaxed.
For instance, relaxing the closed condition of the $B$-field (\ref{B-closed}) non-trivially modifies the field equation for the dilaton (\ref{eom_dilaton-covconst1}),
even under the covariantly constant condition for the scalar fields (\ref{covconst-scalar}).
This deformation would provide different values of the scalar fields on the black hole solution from their vacuum expectation values.
Thus it is quite interesting to import the technologies in \cite{Cacciatori:2009iz, Dall'Agata:2010gj, Hristov:2010ri} to our system.

Introducing D-branes wrapped on (subspaces of) the internal space \cite{Caviezel:2008ik} would be an admissible procedure to construct a charged black hole in flux compactification scenarios.
If D-branes are appropriately wrapped on certain cycles in the internal space,
they would behave as charged particles.
Different from the case of Calabi-Yau compactifications with D-branes,
such the charged particles indeed interact with each other in gauged supergravity.
This implies that the interaction terms in the equations of motion for gauge fields have to be turned on.
Hence the covariantly constant condition must be removed.


\section*{Acknowledgements}

The author would like to thank 
Takahiro Nishinaka for collaboration in the early stage of this project.
He is also grateful to 
Tsuguhiko Asakawa,
Davide Cassani,
Jan Louis,
Masaki Shigemori,
Paul Smyth,
Stefan Vandoren,
and Oscar Varela
for valuable discussions.
He would also like to thank Universit\"{a}t Hamburg/DESY Theory Group for the hospitality during his stay.
This work was supported in part by the JSPS Institutional Program for
Young Researcher Overseas Visits (\#R54).

\begin{appendix}



\section{Convention and profile of the compactification on $G_2/SU(3)$}
\label{app-conventions}

Here we exhibit the minor differences of the convention among this article and the two papers \cite{Cassani:2008rb, Cassani:2009na}:
\begin{center}
\slb{1}{\begin{tabular}{c||c|c|c} \hline
& this article & the paper \cite{Cassani:2008rb} & the paper \cite{Cassani:2009na} \\ \hline\hline
symplectic vectors & $\Pi_{\text{H}}$, $\Pi_{\text{V}}$ & --- & $\Pi_1$, $\Pi_2$ \\
K\"{a}hler potential & $K_{\text{V}}$ & $K_+$ & $K_2$ \\
RR flux charges & $(m^{\Lambda}_{\text{R}}, e_{\text{R} \Lambda})$ & $(m^A_{\text{R}}, e_{\text{R}A})$ & $(p^A, q_A)$ \\
geometric flux charges & $(e_{\Lambda}{}^I, e_{\Lambda I})$ & $(m_{A}{}^I, e_{A I})$ & $(e_{A}{}^I, e_{A I})$ 
 \\ \hline
\end{tabular}}
\end{center}


Here let us briefly summarize the profile of the compactification on the coset space $G_2/SU(3)$.
The explicit form of the period matrix $\N_{\Lambda \Sigma}$ given by the prepotential (\ref{TTT-info-NSR1}) is 
\bsubeq \label{TTT-periodN-NSR1}
\begin{align}
\nu_{\Lambda \Sigma} (\t, \ol{\t}) \ \equiv \ 
\Re\N_{\Lambda \Sigma} \ &= \ 
\frac{\mathscr{I} (\t + \ol{\t})}{4} \left(
\begin{array}{cc}
(\t + \ol{\t})^2 & - 3 (\t + \ol{\t}) \\
- 3 (\t + \ol{\t}) & 12 
\end{array} \right)
\, , \\
\mu_{\Lambda \Sigma} (\t, \ol{\t}) \ \equiv \ 
\Im\N_{\Lambda \Sigma} \ &= \ 
- \frac{\i \mathscr{I}(\t - \ol{\t})}{4} \left(
\begin{array}{cc}
\t^2 + 4 \t \ol{\t} + \ol{\t}{}^2 & - 3 (\t + \ol{\t}) \\
- 3 (\t + \ol{\t}) & 6
\end{array} \right)
\, .
\end{align}
\esubeq
There are three AdS vacua of the system (\ref{actionNSR1}) studied in \cite{Cassani:2009ck}:
\bsubeq 
\begin{gather}
\begin{array}{rcl@{\ls}rcl}
\t_* \!\!&=&\!\!
\dps 
\frac{\pm 1 - \i \sqrt{15}}{2}
\left(
\frac{3}{5 (e_{10})^2} 
\left| \frac{e_{\text{R}0}}{m_{\text{R}}^0} \right|
\right)^{1/3} 
\, , &
\xi^0_* \!\!&=&\!\!
\dps 
- \frac{2}{5} \left( 
\frac{2 \sqrt{3} m_{\text{R}}^0 (e_{\text{R}0})^2}{5 e_{10}} 
\right)^{1/3}
\, , \\ 
\exp ({\varphi}_*) \!\!&=&\!\!
\dps
\frac{4}{3} 
\left(
\frac{\sqrt{5} e_{10}}{\sqrt{3} m_{\text{R}}^0 (e_{\text{R}0})^2} 
\right)^{1/3} 
\, , &
V_* \!\!&=&\!\!
\dps 
- \frac{5 \sqrt{5}}{2}
\left( 
\frac{5 (e_{10})^4}{2 \sqrt{3} | m_{\text{R}}^0 (e_{\text{R}0})^5 | }  
\right)^{1/3}
\, ,
\end{array}
\label{sol-NSR1-vacua1} \\
\begin{array}{rcl@{\ls}rcl}
\t_* \!\!&=&\!\!
\dps 
\big( \pm 1 - \i \sqrt{3} \big)
\left( 
\frac{3}{5 (e_{10})^2} 
\left| \frac{e_{\text{R}0}}{m_{\text{R}}^0} \right|
\right)^{1/3} 
\, , &
\xi^0_* \!\!&=&\!\!
\dps \left(
\frac{9 \, m_{\text{R}}^0 (e_{\text{R}0})^2}{25 e_{10}} 
\right)^{1/3}
\, , \\
\exp ({\varphi}_*) \!\!&=&\!\!
\dps \frac{2}{3} 
\left( 
\frac{25 e_{10}}{\sqrt{3} m_{\text{R}}^0 (e_{\text{R}0})^2} 
\right)^{1/3} 
\, , &
V_* \!\!&=&\!\!
\dps 
- \frac{80}{27}
\left(
\frac{25 (e_{10})^4}{\sqrt{3} | m_{\text{R}}^0 (e_{\text{R}0})^5 | } 
\right)^{1/3}
\, .
\end{array}
\label{sol-NSR1-vacua2} \\
\begin{array}{rcl@{\ls}rcl}
\t_* \!\!&=&\!\!
\dps 
- \i \left( 
\frac{12}{\sqrt{5} (e_{10})^2} 
\left| \frac{e_{\text{R}0}}{m_{\text{R}}^0} \right|
\right)^{1/3} 
\, , &
\xi^0_* \!\!&=&\!\!
0
\, , \\
\exp ({\varphi}_*) \!\!&=&\!\!
\dps \sqrt{5}
\left( 
\frac{5 e_{10}}{18 m_{\text{R}}^0 (e_{\text{R}0})^2} 
\right)^{1/3} 
\, , &
V_* \!\!&=&\!\!
\dps
- \frac{25 \sqrt{5}}{6}
\left( 
\frac{5 (e_{10})^4}{18 | m_{\text{R}}^0 (e_{\text{R}0})^5 | } 
\right)^{1/3}
\, . 
\end{array}
\label{sol-NSR1-vacua3}
\end{gather}
\esubeq
It turns out that the Romans' mass parameter $m_{\text{R}}^0$ must be positive in order that the exponential value of the dilaton.
It is recognized that the vacuum given by (\ref{sol-NSR1-vacua1}) has $\N=1$ supersymmetry, whilst the other two vacua described by (\ref{sol-NSR1-vacua2}) and (\ref{sol-NSR1-vacua3}) are non-supersymmetric \cite{Cassani:2009ck}.
Notice that the scalar field $\wt{\xi}_0$ is not fixed because this does not contribute to the scalar potential.

If one goes back to a Calabi-Yau compactification,
one has to take the vanishing limit of $m_{\text{R}}^0$, $e_{\text{R}0}$ and $e_{10}$, whilst their power orders are different: ${\cal O}(m_{\text{R}0}) = {\cal O}(m_{\text{R}}^0)$ and ${\cal O}(e_{10}) = {\cal O}((e_{\text{R}0})^3)$. 
Under this limit the scalar potential is driven to zero, even though the scalars $\t$ and $\xi^0$ are finite.

\end{appendix}

}

\begin{thebibliography}{99}

\bibitem{Grana:2005jc}
  M.~Gra\~{n}a,
  ``{\sl Flux compactifications in string theory: A comprehensive review},''
  Phys.\ Rept.\  {\bf 423} (2006) 91
  [{arXiv:hep-th/0509003}].

\bibitem{Grana:2005ny}
  M.~Gra\~{n}a, J.~Louis and D.~Waldram,
  ``{\sl Hitchin functionals in $\N=2$ supergravity},''
  JHEP {\bf 0601} (2006) 008
  [arXiv:hep-th/0505264].

\bibitem{Grana:2006hr}
  M.~Gra\~{n}a, J.~Louis and D.~Waldram,
  ``{\sl $SU(3) \times SU(3)$ compactification and mirror duals of magnetic fluxes},''
  JHEP {\bf 0704} (2007) 101
  [arXiv:hep-th/0612237].

\bibitem{D'Auria:2007ay}
  R.~D'Auria, S.~Ferrara and M.~Trigiante,
  ``{\sl On the supergravity formulation of mirror symmetry in generalized
  Calabi-Yau manifolds},''
  Nucl.\ Phys.\  B {\bf 780} (2007) 28
  [arXiv:hep-th/0701247].

\bibitem{Cassani:2008rb}
  D.~Cassani,
  ``{\sl Reducing democratic type II supergravity on $SU(3) \times SU(3)$ structures},''
  JHEP {\bf 0806} (2008) 027
  [{arXiv:0804.0595 [hep-th]}].



\bibitem{Andrianopoli:1996cm}
  L. Andrianopoli, M. Bertolini, A. Ceresole, R. D'Auria, S. Ferrara, P. Fr\'{e} and T. Magri,
  ``{\sl ${\N} = 2$ supergravity and ${\N} = 2$ super Yang-Mills theory on general scalar manifolds: Symplectic covariance, gaugings and the momentum map},''
  J. Geom. Phys. {\bf 23} (1997) 111
[{arXiv:hep-th/9605032}].

\bibitem{Dall'Agata:2003yr}
  G.~Dall'Agata, R.~D'Auria, L.~Sommovigo and S.~Vaul\`{a},
  ``{\sl $D = 4$, $\N=2$ gauged supergravity in the presence of tensor multiplets},''
  Nucl.\ Phys.\  B {\bf 682} (2004) 243
  [{arXiv:hep-th/0312210}].

\bibitem{D'Auria:2004yi}
  R.~D'Auria, L.~Sommovigo and S.~Vaul\`{a},
  ``{\sl $\N = 2$ supergravity Lagrangian coupled to tensor multiplets with electric and magnetic fluxes},''
  JHEP {\bf 0411} (2004) 028
  [{arXiv:hep-th/0409097}].

\bibitem{Cassani:2009na}
  D.~Cassani, S.~Ferrara, A.~Marrani, J.~F.~Morales and H.~Samtleben,
  ``{\sl A special road to AdS vacua},''
  JHEP {\bf 1002} (2010) 027
  [{arXiv:0911.2708 [hep-th]}].



\bibitem{Lust:2004ig}
  D.~L\"{u}st and D.~Tsimpis,
  ``{\sl Supersymmetric AdS$_4$ compactifications of IIA supergravity},''
  JHEP {\bf 0502} (2005) 027
  [arXiv:hep-th/0412250].

\bibitem{Romans:1991nq}
  L.~J.~Romans,
  ``{\sl Supersymmetric, cold and lukewarm black holes in cosmological
  Einstein-Maxwell theory},''
  Nucl.\ Phys.\  B {\bf 383} (1992) 395
  [arXiv:hep-th/9203018].

\bibitem{Caldarelli:1998hg}
  M.~M.~Caldarelli and D.~Klemm,
  ``{\sl Supersymmetry of Anti-de Sitter black holes},''
  Nucl.\ Phys.\  B {\bf 545} (1999) 434
  [{arXiv:hep-th/9808097}].

\bibitem{Sabra:1999ux}
  W.~A.~Sabra,
  ``{\sl Anti-de Sitter BPS black holes in $\N=2$ gauged supergravity},''
  Phys.\ Lett.\  B {\bf 458} (1999) 36
  [{arXiv:hep-th/9903143}].

\bibitem{Chamseddine:2000bk}
  A.~H.~Chamseddine and W.~A.~Sabra,
  ``{\sl Magnetic and dyonic black holes in $D = 4$ gauged supergravity},''
  Phys.\ Lett.\  B {\bf 485} (2000) 301
  [{arXiv:hep-th/0003213}].

\bibitem{Cacciatori:2009iz}
  S.~L.~Cacciatori and D.~Klemm,
  ``{\sl Supersymmetric AdS$_4$ black holes and attractors},''
  JHEP {\bf 1001} (2010) 085
  [{arXiv:0911.4926 [hep-th]}].

\bibitem{Dall'Agata:2010gj}
  G.~Dall'Agata and A.~Gnecchi,
  ``{\sl Flow equations and attractors for black holes in $\N = 2$ $U(1)$ gauged
  supergravity},''
  JHEP {\bf 1103} (2011) 037
  [{arXiv:1012.3756 [hep-th]}].

\bibitem{Hristov:2010ri}
  K.~Hristov and S.~Vandoren,
  ``{\sl Static supersymmetric black holes in AdS$_4$ with spherical symmetry},''
  JHEP {\bf 1104} (2011) 047
  [{arXiv:1012.4314 [hep-th]}].

\bibitem{Bellucci:2008cb}
  S.~Bellucci, S.~Ferrara, A.~Marrani and A.~Yeranyan,
  ``{\sl $d=4$ black hole attractors in $\N=2$ supergravity with Fayet-Iliopoulos terms},''
  Phys.\ Rev.\  D {\bf 77} (2008) 085027
  [{arXiv:0802.0141 [hep-th]}].


\bibitem{Hristov:2010eu}
  K.~Hristov, H.~Looyestijn and S.~Vandoren,
  ``{\sl BPS black holes in $\N=2$ $D=4$ gauged supergravities},''
  JHEP {\bf 1008} (2010) 103
  [{arXiv:1005.3650 [hep-th]}].

\bibitem{Ferrara:1997tw}
  S.~Ferrara, G.~W.~Gibbons and R.~Kallosh,
  ``{\sl Black holes and critical points in moduli space},''
  Nucl.\ Phys.\  B {\bf 500} (1997) 75
  [{arXiv:hep-th/9702103}].

\bibitem{Kallosh:2006bt}
  R.~Kallosh, N.~Sivanandam and M.~Soroush,
  ``{\sl The non-BPS black hole attractor equation},''
  JHEP {\bf 0603} (2006) 060
  [{arXiv:hep-th/0602005}].

\bibitem{Kallosh:2006ib}
  R.~Kallosh, N.~Sivanandam and M.~Soroush,
  ``{\sl Exact attractive non-BPS STU black holes},''
  Phys.\ Rev.\  D {\bf 74} (2006) 065008
  [{arXiv:hep-th/0606263}].

\bibitem{Nampuri:2007gv}
  S.~Nampuri, P.~K.~Tripathy and S.~P.~Trivedi,
  ``{\sl On the stability of non-supersymmetric attractors in string theory},''
  JHEP {\bf 0708} (2007) 054
  [{arXiv:0705.4554 [hep-th]}].

\bibitem{Bellucci:2006zz}
  S.~Bellucci, S.~Ferrara and A.~Marrani,
  ``{\sl Supersymmetric mechanics. Vol. 2: The attractor mechanism and space time singularities},''
  Lect.\ Notes Phys.\  {\bf 701} (2006) 1.

\bibitem{Louis:2002ny}
  J.~Louis and A.~Micu,
  ``{\sl Type II theories compactified on Calabi-Yau threefolds in the presence of background fluxes},''
  Nucl.\ Phys.\  B {\bf 635} (2002) 395
  [arXiv:hep-th/0202168].

\bibitem{Sommovigo:2004vj}
  L.~Sommovigo and S.~Vaul\`{a},
  ``{\sl $D=4$, $\N=2$ supergravity with Abelian electric and magnetic charge},''
  Phys.\ Lett.\  B {\bf 602} (2004) 130
  [{arXiv:hep-th/0407205}].

\bibitem{D'Auria:2004tr}
  R.~D'Auria, S.~Ferrara, M.~Trigiante and S.~Vaul\`{a},
  ``{\sl Gauging the Heisenberg algebra of special quaternionic manifolds},''
  Phys.\ Lett.\  B {\bf 610} (2005) 147
  [{arXiv:hep-th/0410290}].

\bibitem{KashaniPoor:2007tr}
  A.~K.~Kashani-Poor,
  ``{\sl Nearly K\"{a}hler reduction},''
  JHEP {\bf 0711} (2007) 026
  [{arXiv:0709.4482 [hep-th]}].

\bibitem{Cassani:2009ck}
  D.~Cassani and A.~K.~Kashani-Poor,
  ``{\sl Exploiting $\N=2$ in consistent coset reductions of type IIA},''
  Nucl.\ Phys.\  B {\bf 817} (2009) 25
  [{arXiv:0901.4251 [hep-th]}].

\bibitem{Caviezel:2008ik}
  C.~Caviezel, P.~Koerber, S.~K\"{o}rs, D.~L\"{u}st, D.~Tsimpis and M.~Zagermann,
  ``{\sl The effective theory of type IIA AdS$_4$ compactifications on nilmanifolds and cosets},''
  Class.\ Quant.\ Grav.\  {\bf 26} (2009) 025014
  [arXiv:0806.3458 [hep-th]].



\end{thebibliography}
\end{document}